\documentclass[prb,twocolumn,showpacs,preprintnumbers,amsmath,amssymb]{revtex4}

\usepackage[]{graphicx}			% Include figure files
\usepackage{bm}				% bold math
\usepackage{times}				% Times font
\newcommand{\za}{\ensuremath{\alpha}}
\newcommand{\zb}{\ensuremath{\beta}}
\newcommand{\zg}{\ensuremath{\gamma}}
\newcommand{\zd}{\ensuremath{\delta}}

\newcommand{\zz}{\ensuremath{\zeta}}

\newcommand{\zq}{\ensuremath{\theta}}

\newcommand{\zk}{\ensuremath{\kappa}}

\newcommand{\zm}{\ensuremath{\mu}}

\newcommand{\zs}{\ensuremath{\sigma}}

\newcommand{\zf}{\ensuremath{\phi}}

\newcommand{\zw}{\ensuremath{\omega}}

\newcommand{\zG}{\ensuremath{\Gamma}}
\newcommand{\zD}{\ensuremath{\Delta}}

\newcommand{\bk}{\ensuremath{b_k}}
\newcommand{\Bk}{\ensuremath{B_k}}
\newcommand{\wk}{\ensuremath{\zw_k}}
\newcommand{\Gk}{\ensuremath{\zG_k}}
\newcommand{\kb}{\ensuremath{k_\text{B}}}		% Boltzmann constant
\newcommand{\sqrtq}{\ensuremath{\sqrt{q}}}

\newcommand{\pd}{\ensuremath{\partial}}

\newcommand{\half}{\ensuremath{\frac{1}{2}}}

\begin{document}

%%%%%%%%%%%%%%%%%%%%%%%%%%%%%%%%%%%%%%%%
%
%		Header information
%
%%%%%%%%%%%%%%%%%%%%%%%%%%%%%%%%%%%%%%%%

\title{Stochastic theory of spin-transfer oscillator linewidths}

\author{Joo-Von Kim}
\email{joo-von.kim@ief.u-psud.fr}
\affiliation{Institut d'Electronique Fondamentale, UMR CNRS 8622, Universit\'{e} Paris-Sud, 91405 Orsay cedex, France} 
\date{\today}

%%%%%%%%%%%%%%%%%%%%%%%%%%%%%%%%%%%%%%%%
%
%		Abstract
%
%%%%%%%%%%%%%%%%%%%%%%%%%%%%%%%%%%%%%%%%
\begin{abstract}
We present a stochastic theory of linewidths for magnetization oscillations in spin-valve structures driven by spin-polarized currents. Starting from a nonlinear oscillator model derived from spin-wave theory,  we derive Langevin equations for amplitude and phase fluctuations due to the presence of thermal noise. We find that the spectral linewidths are inversely proportional to the spin-wave intensities with a lower bound that is determined purely by modulations in the oscillation frequencies. Reasonable quantitative agreement with recent experimental results from spin-valve nanopillars is demonstrated.
\end{abstract}

%%%		PACS, the Physics and Astronomy Classification Scheme.
%%%		Use showkeys class option if keyword display desired

\maketitle

%%%%%%%%%%%%%%%%%%%%%%%%%%%%%%%%%%%%%%%%
%
%                Paper
%
%%%%%%%%%%%%%%%%%%%%%%%%%%%%%%%%%%%%%%%%

%%%
%%
%	Introduction
%%
%%%
\section{Introduction}

Since the theoretical predictions of Berger~\cite{Berger:PRB:1996} and Slonczewski~\cite{Slonczewski:JMMM:1996, Slonczewski:JMMM:1999} concerning the influence of a spin-polarized current on magnetization at a mesoscopic scale, many issues concerning the interplay between spin transport and spin dynamics in metallic heterostructures have been brought to light by extensive experimental and theoretical work. The primary effect on these length scales is a transfer of spin angular momentum between the current and the magnetization, which may lead to magnetization reversal, the generation of spin waves, or both. One can account for the spin-polarized current in the equations of motion of magnetization $\vec{M}$, to a reasonable approximation, by an additional torque of the form
\begin{equation}
\frac{\pd \vec{M}}{\pd t}\bigg|_{I} = \frac{\zs(\vec{p},\vec{M}) I}{M_s} \vec{M}\times (\vec{M} \times \vec{p}),
\label{eq:stt}
\end{equation}
where $I$ is the current density, $M_s$ is the saturation magnetization, $\vec{p}$ is the unit vector along the direction of spin-polarization, and $\zs$ is a parameter that measures the spin-transfer efficiency. For  particular orientations of magnetization, current polarization and applied fields, such a term can act as a negative Landau-Lifshitz form of magnetic damping. As such, the existence of stable precessional states of magnetization can be made possible by tuning the applied current such that it compensates the intrinsic magnetic damping (averaged over a precession period). These oscillations can be detected experimentally through the oscillating voltages arising from the time-varying magnetoresistance in a spin-valve structure. Evidence for such dynamical states have been reported for a number of geometries and magnetic materials,~\cite{Tsoi:PRL:1998, Tsoi:PRL:2002, Rippard:APL:2003, Kiselev:Nature:2003, Rippard:PRL:2004, Tsoi:PRB:2004, Kiselev:PRL:2004, Covington:PRB:2004, Tsoi:PRB:2004b, Tsoi:PRL:2004, Rippard:PRB:2004, Krivorotov:Science:2005, Kiselev:PRB:2005, Kaka:Nature:2005, Mancoff:Nature:2005, Mistral:MSEB:2006} which seems to indicate that such stable oscillations are a general feature of spin-momentum transfer.

While the frequencies of these oscillations appear to be well-understood from spin-wave theory~\cite{Rezende:PRL:2005,Slavin:IEEE:2005} and numerical simulation,~\cite{Lee:NM:2004,  Xi:PRB:2004, Russek:PRB:2005, Bertotti:PRL:2005, Montigny:JAP:2005, Xiao:PRB:2005} there is less consensus concerning the origin of the spectral linewidths. Part of the reason for this discrepancy may be due to the different experimental conditions, materials, and heterostructure geometries used to study these oscillations. For instance, it is found that spectral linewidths for point-contact geometries~\cite{Rippard:PRL:2004,Rippard:PRB:2004, Kaka:Nature:2005} are generally narrower than those reported for nanopillars,~\cite{Kiselev:Nature:2003, Kiselev:PRB:2005} although recent experimental evidence from more complex nanopillar stacks indicates that this is not a general rule.~\cite{Krivorotov:Science:2005, Mistral:2006} Moreover, simple macrospin calculations~\cite{Russek:PRB:2005} and more elaborate micromagnetics simulations~\cite{Lee:NM:2004,Montigny:JAP:2005} consistently overestimate the spectral linewidths by at least an order of magnitude.

In this article, we present a stochastic theory of spin-transfer oscillator linewidths based on spin-wave theory. We derive a linearized Langevin equation to show that fluctuations about the dynamic state, due to thermal noise, gives rise to a spectral linewidth that is inversely proportional to the spin-wave mode intensity with a lower bound determined by modulations in the oscillation frequency. The main features of the theory are shown to give reasonable quantitative agreement with recent experiments on spin-valve nanopillars.

%%%
%%
%	Introduction
%%
%%%
\section{Stochastic model}

We believe that the current-induced magnetization oscillations observed in experiment correspond to parametrically excited spin-wave modes. Parametric excitation here differs from more well-known means such as parallel pumping,~\cite{Morgenthaler:JAP:1960,Schlomann:JAP:1960} in which parametric resonance is achieved with an oscillating magnetic field applied parallel to the magnetization. In the case of spin-transfer, the torque in (\ref{eq:stt}) allows a means of modifying the spin-wave damping, so parametric excitation occurs as the resultant damping for a given spin-wave mode becomes negative. The exact profiles of the excited modes are geometry-dependent, but nevertheless one can derive from spin-wave theory~\cite{Lvov:1994,Slavin:IEEE:2005,Rezende:PRL:2005} a generic oscillator equation of the form (here we follow the notation of Ref.~\onlinecite{Slavin:IEEE:2005}),
\begin{equation}
\dot{b}_k + (i \zw_{k0} + \zG_k) \bk = T_k(\{b\},H,I,\dots) + f_k(t),
\label{eq:sto_generic}
\end{equation}
where $\bk(t)$ is a complex dimensionless variable ($|b|<1$) for a spin-wave mode $k$, $\zw_{k0}$ the spin-wave frequency, and $\zG_k$ is a phenomenological relaxation rate that encompasses all possible damping mechanisms for the spin wave. All information concerning the relevant magnetic energies (Zeeman, exchange, magnetocrystalline anisotropy, dipole-dipole, etc.) are contained in $\zw_{k0}$ through the dispersion relation for the spin waves. The $b_k$ variables diagonalize the linear part of the original magnetic Hamiltonian $\mathcal{H}_0$ (without current terms) and can be found from the usual series of Holstein-Primakoff and Bogoliubov transformations (a discussion pertinent to thin films can be found in Ref.~\onlinecite{Slavin:1994}). For finite magnetic elements of arbitrary shape, it is possible to find the corresponding eigenmodes numerically~\cite{Giovannini:PRB:2004,Grimsditch:PRB:2004, McMichael:JAP:2005} or through the use of approximate spin-wave theories.~\cite{Guslienko:PRB:2003,Guslienko:PRB:2005,Bayer:PRB:2005} The quantity $T_k$ on the right-hand side represents nonlinear terms that include higher-order interactions and sources of parametric excitations, such as Eq.~\ref{eq:stt}. We assume that the experimentally observed oscillation corresponds to a parametrically excited spin-wave mode $k$ that is largely populated in comparison to the thermal occupation of other modes. This mode is taken to be immersed in a thermal bath whose effect is represented by the additional stochastic force $f_k(t)$ acting on $\bk$, where $f_k(t)$ is taken to represent a white noise associated with thermal magnons,
\begin{equation}
\langle f_k(t) f_{k'}^{*}(t')\rangle = 2 \Gk \frac{n_0(\wk)}{NS} \zd_{k,k'} \zd(t-t').
\label{eq:thermalmagnons}
\end{equation}
$n_0(\wk)$ is the thermal occupation of magnons and is normalized by $NS = M_s V/(g \mu_B)$, the total spin in the system, in the same manner as the variables $\bk$. Here, $M_s$ is the saturation magnetization, $V$ the volume of the magnetic element, $g$ the gyromagnetic factor, and $\mu_B$ the Bohr magneton.

In the absence of the nonlinear $T_k$ term, it is straightforward to obtain from direct integration of Eq. \ref{eq:sto_generic} 
\begin{equation}
\half \frac{d n_k}{d t} = -\Gk (n_k - n'_0).
\label{eq:magnonpop}
\end{equation}
Thus, fluctuations in the magnon population  $n_k = \langle \bk^{*} \bk \rangle$ relax towards $n'_0 \equiv n_0(\wk)/(NS)$ with a characteristic time $\Gk^{-1}$. At thermal equilibrium, in the absence of driving terms, the occupation of thermal magnons must go over to the usual Bose-Einstein distribution, $n_0 = (\exp[\hbar \wk/\kb T] - 1)^{-1}$. For spin-waves in the GHz range, it suffices to use $n_0 = \kb T/ (\hbar \wk)$. In what follows, we shall assume that this relationship holds for all levels of driving and nonlinearity considered here. We therefore neglect the role of nonlinear damping on the thermal noise of the system~\cite{Mikhailov:JETP:1976}; appropriate estimates of such higher-order effects, obtained for parallel pumping, are given in Ref.~\onlinecite{Zakharov:JETP:1971}.

As discussed above, a sufficiently large spin-polarized current can result in a stable dynamic oscillation of magnetization. By expanding Eq.~\ref{eq:stt} in the appropriate spin-wave variables, neglecting all three-wave interaction terms and retaining only four-wave terms that lead to self-interaction of the single excited spin-wave mode, one finds an oscillator equation with the additional stochastic term of the form~\cite{Slavin:IEEE:2005}
\begin{equation}
\dot{b}_k + (i \wk + \zG_k) \bk = \zs I (1 - |\bk|^2)\bk + f_k(t),
\label{eq:sto_stochastic}
\end{equation}
where $\wk = \zw_{k0} + N_k |\bk|^2$ contains a nonlinear frequency shift that can become important for large spin-wave amplitudes. In the absence of the stochastic term, Eq.~\ref{eq:sto_stochastic} describes the dynamics of a nonlinear oscillator that is capable of supporting self-sustained oscillations with a frequency $\wk$. Indeed, at the threshold $\zs I = \zG_k$ with $f_k=0$, a Hopf bifurcation takes place and a stable orbit with $|\bk|^2 = d \equiv (\zs I - \Gk)/\zs I $ is made possible for $\zs I > \zG_k$.

\begin{figure}
\includegraphics[width=7cm]{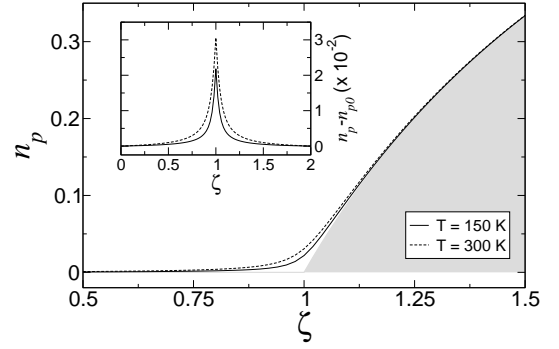}
\caption{\label{fig:nk_v_zeta}Thermal dependence of parametric magnon population $n_p$, as a function of supercriticality $\zz$. The shaded region corresponds to the zero temperature case $n_{p0}$. The difference between finite- and zero-temperature populations is shown in the inset. }
\end{figure}

It is interesting to note that the presence of thermal noise suppresses completely the current threshold for self-sustained oscillations. To see this, we introduce the nonlinear terms in (\ref{eq:sto_stochastic}) into (\ref{eq:magnonpop}) and take the steady-state limit to obtain
\begin{equation}
\zz n_k^2 + (1-\zz)n_k - n'_0 = 0,
\label{eq:thermpara}
\end{equation}
where $\zz \equiv \zs I/\Gk$ is the zero-temperature supercriticality parameter. It is straightforward to show that the presence of a non-zero $n'_0$ term leads to a finite parametric magnon population $n_p = n_k - n'_0$ for all currents $I>0$, as shown in Fig.~\ref{fig:nk_v_zeta}. 
For sufficiently large currents such that $\zz$ is far from 1, we see that the is very little difference between the zero-temperature and finite-temperature parametric magnon populations. As such, we will assume that $|\bk|^2 = d$ for self-sustained oscillations holds at all temperatures to simplify the proceeding analysis, and use of the actual temperature dependence will be made later when discussing linewidths.

In addition to the stochastic force $f_k(t)$, noise sources can lead to a direct modulation of oscillation frequency $\wk$ itself. Such modulations can originate from external field sources or other fluctuating magnetic elements in close proximity. An estimate of the frequency modulations~\cite{Kubo:1962} can be made by considering the action of a random thermal field $h_r(t)$ superimposed on the external field $H_0$, $H(t) = H_0 + h_r(t)$, where $h_r(t)$ is assumed to be a white noise with zero mean,
\begin{equation}
\langle h_r(t) \rangle = 0;  \langle h_r(t) h_r(t')\rangle = 2 D_r \zd(t-t'), 
\end{equation}
and $D_r$ is an appropriate diffusion constant. The action of the random field on the equations of motion for the $b$ variables are obtained by supplementing the original diagonal Hamiltonian $\mathcal{H}_0$ with the random field Hamiltonian
\begin{equation}
\mathcal{H}_r(t) = -\mu \sum_i h_{r}(t) S_{iz} \simeq \hbar \zw_r(t) \sum_k  \bk^{*}\bk,
\end{equation}
where $\mu = g \mu_0 \mu_B$, $\zw_r(t) = \zg h_r(t)$, and $\zg = g \mu_0 \mu_B / \hbar$ is the gyromagnetic constant. Following the usual procedure for generating the equations of motion, $\dot{b}_k = i [\mathcal{H}_0 + \mathcal{H}_r,\bk]/\hbar$, it is a simple matter to show that the random field leads to a modulation of the spin-wave frequency $\zw(t) = \zw_{0} + \zw'(t)$.

In experiment, the current-driven oscillations in spin-valve devices are detected through a time-varying voltage due to the corresponding magnetoresistance oscillations. Giant magnetoresistance is related to the relative orientations of the free $\vec{m}$ and fixed magnetizations $\vec{M} = M \hat{z}$, $R = R_0  + \zD R \; (\vec{M} \cdot \vec{m})$, where $R_0$ and $\zD R$ are material-dependent constants that  can be readily determined in experiment. We allow for a misalignment between the easy axes of the free and reference layers, which is denoted by the angle $\zq$ measured from the $z$ axis in the $xz$ plane. The corresponding (scaled) voltage fluctuations are then given by
\begin{equation}
v \equiv \frac{V-I R_0}{I \zD R} =  m_z \cos{\zq} - \half \sin{\zq} (m_{+} + m_{-}),
\end{equation}
where $m_{\pm} = m_x \pm i m_y$, and $I_0$ is the average dc current. Expanding $\vec{m}$ in the appropriate spin-wave variables and assuming that only the parametrically excited mode $k$ contributes to the time-varying magnetization, we find that the two-time correlation function for the voltage fluctuations, $\langle v(t) v(0) \rangle$, can be written as
\begin{multline}
\langle v(t) v(0) \rangle = F(k,\zq) ( \langle b_k(t) b_k^{*}(0) \rangle +  \text{ c.c.}) \\
+ G(k,\zq) \langle b_k(t) b_k^{*}(t) b_k(0) b_k^{*}(0) \rangle ...,
\label{eq:voltage}
\end{multline}
where $F(k,\zq)$ and $G(k,\zq)$ contain all the factors related to the transformation to spin-wave variables. To obtain (\ref{eq:voltage}), we have assumed that there exists a center of symmetry such that $\bk = b_{-k}$ and that the spatial average of $\bk$ is non-zero. These assumptions are justified in rectangular magnetic elements, for example, in which the lowest-lying excitations have been shown to be cosinusoidal standing waves in the dipolar-dominated regime. In the absence of misalignment ($\zq = 0$), contributions to the magnetoresistance oscillations can only come from the $m_z$ term, which gives rise to the $G(k,\zq)$ terms to lowest order and  result in spectral frequencies that are twice the oscillation frequencies. In what follows, we assume that the misalignment is non-zero and that the two-time correlation functions in the spin-wave variables, proportional to $F(k,\zq)$, are sufficient to describe the experimental data.

%%%
%%
%	Correlation functions
%%
%%%
\section{Correlation functions}

The Fourier transform of $\langle v(t) v(0) \rangle$ gives the spectral density of voltage oscillations measured in experiment, so the main goal of this section is to obtain expressions for $\langle b_k(t) b_k^{*}(0) \rangle$ and  $\langle b_k^{*}(t) b_k(0) \rangle$ from (\ref{eq:sto_stochastic}). To begin, we make the assumption that fluctuations in the spin-wave amplitudes are sufficiently small so as not to modify the frequency in the stable oscillation regime. This allows a rotating-wave approximation [$\bk(t) = \Bk(t) e^{-i \zw t}$] to be made to obtain
\begin{equation}
\dot{B}_k =  \zb (d_k - |\Bk|^2) \Bk + \sqrt{q} \tilde{f}_k(t),
\label{eq:rotwaveapprox}
\end{equation}
where $\zb = \zs I$, $d_k = (\zs I - \Gk)/\zb$, $\sqrt{q} \tilde{f}_k(t) = e^{i \zw t} f_k(t)$, and $q = \Gk n_0/2NS$. It is reasonable to assume that the stochastic noise remains white in the rotating frame, so we will take $\tilde{f}_k(t)$ to have zero mean and satisfy  $\langle \tilde{f}_k(t) \tilde{f}^{*}_{k'}(t') \rangle = 4 \delta_{k,k'}\delta(t-t')$. For the sake of clarity and without loss of generality, we will omit the mode index $k$ in what follows. Following the treatment in Ref.~\onlinecite{Risken:1989}, we introduce $B = b_1 + i b_2$ and $\tilde{f} = f_1 + i f_2$ to write Eq. \ref{eq:rotwaveapprox} as
\begin{equation}
\dot{b}_i = \zb (d-b_1^2 - b_2^2)b_i + \sqrt{q} f_i(t),
\end{equation}
where $i=1,2$ and $\langle f_i(t) f_j(t) \rangle = 2 \delta_{i,j} \delta(t-t')$. Because we are interested in separating out the amplitude and phase fluctuations in $B$, we find it convenient to introduce $B = A \exp(i \zf)$, where $A = \sqrt{b_1^2 + b_2^2}$ and $\zf = \tan^{-1}(b_2/b_1)$, which leads to 
\begin{align}
&\dot{A} - \zb (d-A^2)A = \sqrtq (f_1 \cos{\zf} + f_2 \sin{\zf}); \notag \\
&\dot{\zf} = \frac{\sqrtq}{A} (-f_1 \sin{\zf} + f_2 \cos{\zf}). \label{eq:rotwave1} 
\end{align}
Instead of solving (\ref{eq:rotwave1}), one can equivalently consider 
\begin{align}
&\dot{A} - \zb (d-A^2)A - \frac{q}{A} = \sqrtq f_A{(t)}; \notag \\
&\dot{\zf} = -\frac{\sqrtq}{A} f_{\zf}(t), \label{eq:rotwave2} 
\end{align}
with
\begin{align}
&\langle f_A (t) f_A (t') \rangle = \langle f_\zf (t) f_\zf (t') \rangle = 2 \delta(t-t'), \notag \\
&\langle f_A (t) f_\zf (t') \rangle = 0,
\label{eq:nonlinear}
\end{align}
as these Langevin equations lead to the same drift and diffusion coefficients~\cite{Risken:1989} (in the Stratonovitch definition) and therefore describe the same physical system.

The stochastic equations (\ref{eq:rotwave2}) represent a set of coupled nonlinear Langevin equations. In general, it is difficult to compute correlation functions directly from these equations and one can instead solve the corresponding Fokker-Planck equation. For our spin-transfer oscillators, we note that for excitations sufficiently above threshold (determined in this case by the thermal noise floor), one can linearize the Langevin equations about the stationary state $|B|^2 = d$ and consider only the small fluctuations about this dynamic equilibrium. Let $A(t) = \sqrt{d} + a(t)$, where $a(t)/\sqrt{d} \ll 1$. If we retain only linear terms in $a(t)$ and neglect a term proportional to $q/d$, we find a simple Ornstein-Uhlenbeck process for the amplitude fluctuations and a Wiener process for the phase fluctuations,
\begin{align}
\dot{a} + \zk a &= \sqrtq f_A{(t)}; \notag  \\
\dot{\zf} &= -\sqrt{\frac{q}{d}} f_{\zf}(t), 
\label{eq:linear} 
\end{align}
where $\zk = 2 \beta d = 2(\zs I  - \Gamma)$. In the absence of the stochastic terms, (\ref{eq:linear}) shows that amplitude fluctuations are damped out with a characteristic frequency $\zk$ that increases as the driving current increases (or equivalently, as the system moves further away from threshold). While this can lead to a linewidth broadening with the driving current, it is shown later that it is the phase fluctuations which dominate in the large amplitude regime.

From the formal solutions of (\ref{eq:linear}),
%%
%
%\begin{align}
%a(t) &= a_0 e^{-\zk t} + \sqrtq \int_0^t dt' \; e^{-\zk (t-t')} f_A(t'), \\
%\zf(t) &= \zf_0 - \sqrt{\frac{q}{d}} \int_0^t  dt' \; f_\zf (t').
%\end{align}
% 
%%
it is straightforward to evaluate the desired two-time correlation functions for large times $t,t' \gg 0$ after which initial correlations are unimportant. We find
\begin{align}
\langle a(t) a(t') \rangle &= \frac{q}{\zk} e^{-\kappa |t-t'|}; \notag \\
\langle e^{i \zf(t)} e^{-i \zf(t')} \rangle &= e^{-q |t-t'|/d},
\end{align}
%
%%
%For the phase fluctuations, we need to evaluate a correlation function of the form $\langle e^{i \zf(t)} e^{-i \zf(t')} \rangle$. This can be done by expanding the exponentials in a power series and using the fact that for Gaussian noise the higher-order correlation functions for the stochastic force are given by
%%
%
%\begin{align}
%\langle f(t_1) f(t_2) \cdots f(t_{2n-1}) \rangle &= 0, \notag \\
%\langle f(t_1) f(t_2) \cdots f(t_{2n}) \rangle &= \notag \\ 
%2^n \sum_{P} \delta(t_{i_1} - t_{i_2})\delta(t_{i_3} &- t_{i_4}) \cdots \delta(t_{i_{2n-1}} - t_{i_{2n}}),
%\end{align}
%
%%
%where the sum is over the $P = (2n)!/(2^n !)$ permutations which give a different expression for $\delta(t_{i_1} - t_{i_2}) \cdots \delta(t_{i_{2n-1}} - t_{i_{2n}})$. We find
%%
%
%\begin{equation}
%\langle e^{i \zf(t)} e^{-i \zf(0)} \rangle = e^{-q |t|/d},
%\end{equation}
%
%%
which shows that phase fluctuations are inversely proportional to the mode intensity $d$ and includes a temperature dependence through $q$.

While the last result demonstrates that phase fluctuations in $B$ are inversely proportional to the mode intensity $d$, there exists a lower limit to the phase noise arising from modulations in the oscillation frequency, which are implicitly excluded from the analysis above by virtue of the rotating-wave approximation. An estimate of this additional contribution to the phase noise is found from
\begin{equation}
\langle e^{i \zw(t)} e^{-i \zw(t')} \rangle = e^{i \zw_0 t} e^{-\zg^2 D_r |t-t'|},
\end{equation}
which shows that frequency modulations will lead to an amplitude-independent linewidth in the spectral density that is proportional to $\zg^2 D_r$. If we take a random magnetic field corresponding to superparamagnetic fluctuations of the entire magnetic element (Goldstone mode),~\cite{Brown:PR:1963} we find
\begin{equation}
\zg^2 D_r = \frac{\zg \za}{\mu_0 M_s V}  \kb T,
\label{eq:randomfield}
\end{equation}
where $\za$ is the Gilbert damping constant. We point out that it is also possible to include the effects of spin-shot noise into this definition by an appropriate renormalization of $\za$.~\cite{Foros:PRL:2005} Other fluctuations in the applied current will not be considered here.

The spectral density of magnetoresistance oscillations observed in spin-transfer experiments is proportional to the Fourier transform of the correlation functions $\langle b(t) b^{*}(0) \rangle$ and  $\langle b^{*}(t) b(0) \rangle$. The contribution to positive frequencies is given by $\langle b^{*}(t) b(0) \rangle$, which in terms of our definitions for $A$ and $\zf$ is
\begin{align}
\langle b^{*}(t) b(0) \rangle &= \langle e^{i \zw(t)} e^{-i \zw(0)} \rangle \langle B^{*}(t) B(0) \rangle, \notag \\
&=  \langle e^{i \zw(t)} e^{-i \zw(0)} \rangle   \langle A(t) A(0) \rangle   \langle e^{i \zf(t)} e^{-i \zf(0)}\rangle, \notag \\
&= e^{i \zw_0 t} e^{-(\zg^2 D_r + q/d)|t|} \bigl(d + \frac{q}{\zk}e^{-\zk |t|} \bigr),
\end{align}
where in the second line we have used the fact that amplitude and phase fluctuations are uncorrelated in the linearized equations. Because the correlation function is a sum of two exponential functions, the spectral density will be a sum of two Lorentzians centered about $\zw = \zw_0$, $S(\zw) = d \; S_1(\zw; \zG^{\text{eff}}_1) + (q/\zk) S_2(\zw; \zG^{\text{eff}}_2)$, where $S_i(\zw; \zG)$ is a normalized Lorentzian,
\begin{equation}
S_i(\zw; \zG) = \frac{1}{\pi} \frac{\zG}{(\zw - \zw_0)^2 + \zG^2},
\end{equation}
with $\zG^{\text{eff}}_1 = \zg^2 D_r + q/d$ and $\zG^{\text{eff}}_2 = \zG^{\text{eff}}_1 + \zk$. While the presence of the $q/\zk$ and $\zk$ factors from the amplitude correlation function appear to introduce a complicated $d$ dependence, 
\begin{equation}
\zk =  \half \zG \frac{4d}{1-d}; \; \frac{q}{\zk} = \frac{n'_0(\zw)}{4}\frac{1-d}{d},
\end{equation}
one can show numerically that, in practice, such terms do not significant modify the overall $1/d$ dependence of the effective linewidth of $S(\zw)$, and as such, one can use
\begin{equation}
S(\zw) = \frac{S_0(d)}{(\zw - \zw_0)^2 + (\zg^2 D + \frac{q}{d})^2},
\end{equation}
where
\begin{equation}
S_0(d) = \frac{2}{\zG} \biggl\{ \frac{d^2}{n'_0} + \biggl[\frac{n'_0}{4}\biggr]\biggl[\frac{(1-d)^2}{4d^2 + n'_0 (1-d)}\biggr] \biggr\},
\label{eq:spectral_intensity}
\end{equation}
to a good approximation.

%%%
%%
%	Results and comparison to experiment
%%
%%%
\section{Results and comparison to experiment}

To obtain the full temperature dependence it is necessary to replace $d$ with the correct parametric magnon population at finite temperatures, 
\begin{equation}
n_p(d,T) = \frac{d}{2} + \sqrt{d^2 + 4(1-d) n'_0(T)} - n'_0(T),
\label{eq:paramagnon}
\end{equation}
which is obtained from the solution of (\ref{eq:thermpara}). Thus, the spectral linewidth is given by
\begin{equation}
\zD\zw_k = 2\zg^2 D_r + \zG_k \frac{\kb T}{n_p(T)}.
\label{eq:full_linewidth}
\end{equation}
This is the main result of this paper. All material parameters are encompassed in the spin-wave relaxation frequency $\zG_k$, while the thermal and intensity dependences of the linewidth are contained in $\kb T/ n_p(T)$. If we apply the Gilbert model for spin-wave relaxation and ignore propagating losses, the damping rate $\zG_k$ can be obtained from~\cite{Slavin:IEEE:2005}
\begin{equation}
\zG_k = \za \zw_{k0} \frac{\pd \zw_{k0}}{\pd \zw_H},
\label{eq:Gilbertdamping}
\end{equation}
where $\zw_H$ represents the circular part of the magnetization precession. Thus, the spectral linewidth for a parametrically excited spin-wave mode $\bk$ with Gilbert damping, subject to thermal noise sources (\ref{eq:thermalmagnons}) and (\ref{eq:randomfield}), results in an appealing simple form
\begin{equation}
\zD\zw_k = \frac{2\zg \za}{\mu_0 M_s V}\biggl( 1  + \half \frac{\pd \zw_k}{\pd \zw_H} \frac{1}{d} \biggr) \kb T.
\label{eq:linewidth}
\end{equation}
This result shows that random field fluctuations given by (\ref{eq:randomfield}) play a minor role compared to the thermal noise, since $\pd \zw_k/\pd \zw_H$ is of order unity and $d \ll 1$ for most practical situations. We will neglect the random field term in what follows, but we would like to emphasize that other external field sources not considered here could, in principle, give more significant contributions.

\begin{figure}
\includegraphics[width=7cm]{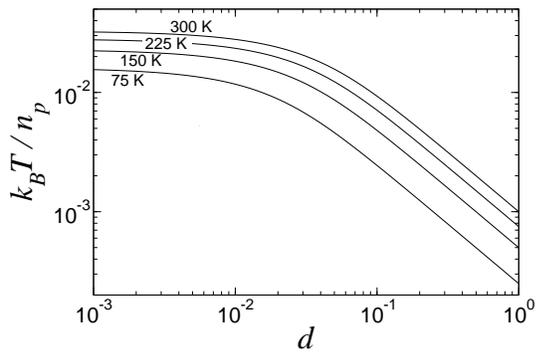}
\caption{\label{fig:linewidth_v_d}Linewidth factor $\kb T/ n_p$ as a function of $d$ for several temperatures.}
\end{figure}

A plot of the linewidth factor $\kb T/ n_p$ as a function of $d$ is presented in Fig.~\ref{fig:linewidth_v_d}.
Here, we have used a parametric spin wave frequency of $\zw_{k0}/(2\pi) =  12$ GHz for the scale factor in the temperature dependence. For low excitation amplitudes, one observes that the linewidth is nearly independent of the intensity $d$ and is instead dominated by the thermal noise term $n'_0$. As parametric excitation is increased past $d \simeq 10^{-2}$, the linewidth decreases rapidly like the inverse of the intensity $1/d$. We would like to point out that such a spin-wave theory is expected to  breakdown somewhere in the region $d > 0.2$, in which it would be necessary to retain higher-order terms in (\ref{eq:sto_stochastic}) for a correct description. Furthermore, there is no guarantee that the single parametric mode will remain stable in this highly nonlinear regime. One should therefore regard the predictions for $d > 0.2$ as an illustration of the general qualitative behavior.

\begin{figure}
\includegraphics[width=7cm]{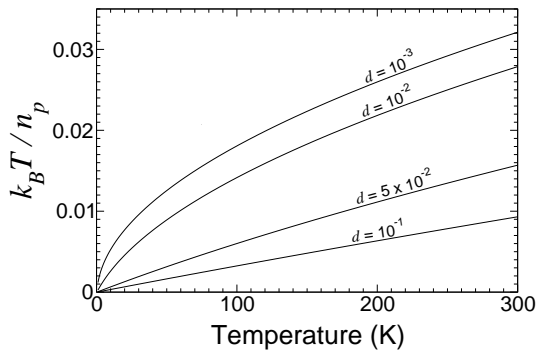}
\caption{\label{fig:linewidth_v_T}Linewidth factor $\kb T/ n_p$ as a function of temperature for several spin-wave intensities $d$.}
\end{figure}

The renormalization of the parametric magnon population by the thermal magnons in (\ref{eq:paramagnon}) leads to a non-trivial thermal dependence for the spectral lines, as shown in Fig.~\ref{fig:linewidth_v_T}.
In the low-amplitude regime where thermal magnons dominate the linewidth, one observes that the linewidth increases like $\sqrt{T}$, due to the corresponding $n'_0$ term under the square-root in (\ref{eq:paramagnon}). Such a behavior has already been observed in experiment~\cite{Sankey:PRB:2005} and can be understood from random-field arguments. As the mode excitation increases, a transition towards a linear temperature dependence occurs between $d=0.01$ and $d=0.1$. In this regime, the thermal magnons are essentially separated from the paramagnetic magnons, so their role is simply to provide an independent thermal noise term with a linear temperature dependence. Such a regime change should be detectable in experiment.

\begin{figure}
\includegraphics[width=8cm]{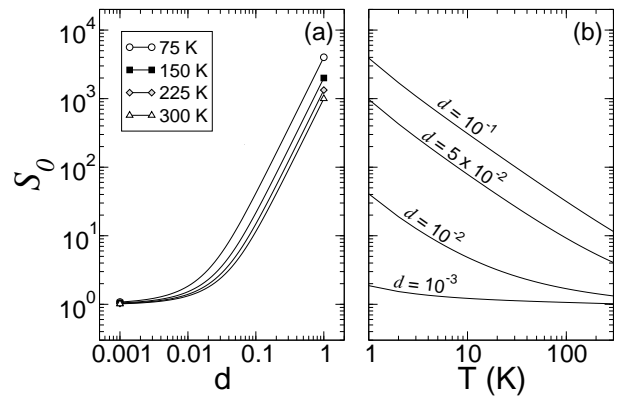}
\caption{\label{fig:line_intensity}Spectral line intensity $S_0(\zw = \zw_{k0})$ as a function of (a) intensity $d$ and (b) temperature $T$. The symbols in (a) are a guide to the eye.}
\end{figure}

Much of the discussion so far has made use of $d$ as a characteristic parameter. Indeed, the results above show that quantitative predictions about the spectral linewidths can be made without knowledge of the spin-transfer efficiency $\zs$ (and therefore the absolute values of the threshold currents), which can only be obtained from detailed transport calculations.~\cite{Xiao:PRB:2004} In the same spirit, we believe that quantitative comparisons to experiment can be made through a characteristic feature of the parametric spin wave spectra that is independent of the current and field configurations. We suggest that the intensity of the spectral lines, given by (\ref{eq:spectral_intensity}), is a good experimental parameter against which linewidths can be compared. In Fig.~\ref{fig:line_intensity}a, the theoretical spectral line intensity $S_0$ in (\ref{eq:spectral_intensity}) is shown as a function of mode intensity $d$.
As one would expect, the power pumped into the parametric mode increases as the mode intensity increases, with a low-amplitude region that is essentially flat due to thermal magnons. Departures from this monotonic increase, observed in any experiment, would be a sign of mode instability and the generation of other spin-wave modes. A stronger temperature dependence is seen for large amplitudes (Fig.~\ref{fig:line_intensity}b).

\begin{figure}
\includegraphics[width=7cm]{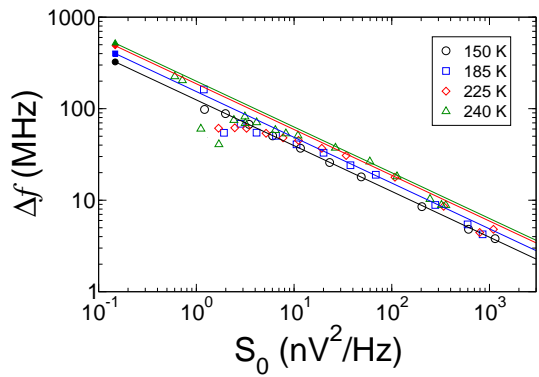}
\caption{\label{fig:linewidth_v_S}(Color online) Comparison of Eq.~\ref{eq:linewidth} (solid lines) with experimental data from Ref.~\onlinecite{Mistral:2006}. The only fitting parameter is the scale factor relating the measured mode intensity $S_0 \propto n_p^2$ to the parametric magnon population $n_p$, which is obtained from one data point from the 150 K series.}
\end{figure}

In the regime $d \gtrsim 0.02$, the spectral line intensity is approximately proportional to the square of the spin-wave intensity $d$. Therefore, one can expect the relationship $\zD\zw_k \propto 1/S_0^2$ to hold for large excitation amplitudes. An example of this behavior has been observed in recent experiments on rectangular spin-valve nanopillars.~\cite{Mistral:2006} The experimental results with a comparison of (\ref{eq:linewidth}) and (\ref{eq:spectral_intensity}) for large $d$ are shown Fig.~\ref{fig:linewidth_v_S}.
The free-layer in the spin-valve stack has dimensions of 100 $\times$ 50 $\times$ 2.8 nm, with $\mu_0 M_s$ = 0.85 T and a Gilbert damping constant of $\za = 0.02$. The data are taken from a low-current regime in which only one oscillation mode is observed and no evidence of mode instability is detected. The solid lines in Fig.~\ref{fig:linewidth_v_S} are based on the theoretical prediction of (\ref{eq:linewidth}), where the smallest linewidth value from the 150 K data was used to obtain a correspondence between the measured spectral line intensity $S_0$ and the parametric magnon population $n_p$; this is the only fitting parameter. To calculate the spin-wave damping rate (Eq.~\ref{eq:Gilbertdamping}), we used the dispersion relation appropriate for rectangular magnetic elements and assumed that the lowest-order spin-wave mode is excited.~\cite{Bayer:PRB:2005} All other parameters are obtained from experiment.~\cite{Mistral:2006} One observes that an excellent fit is obtained for the lowest temperature data set (150 K), with progressively larger deviations as the temperature is increased, the origins of which are still under investigation. Nevertheless, the theoretical curves give a good quantitative bound on the experimental linewidths. We note that the random field contribution (\ref{eq:randomfield}) would lead to a limiting linewidth of about 0.1 MHz with these experimental parameters.

%%%
%%
%	Discussion and concluding remarks
%%
%%%
\section{Discussion and concluding remarks}

The quantitative agreement between our stochastic model and experiment lends credence to the hypothesis of parametric spin-wave excitation in spin-transfer oscillators.~\cite{Rezende:PRL:2005,Slavin:IEEE:2005} This description is most appropriate for nanopillar geometries in which the eigenmode spectrum is well defined. For point-contact geometries, self-localization of spin-wave excitations has been shown to be crucial to describe experimental data.~\cite{Slavin:PRL:2005} One could, in principle, extend the ideas in this article to point-contact geometries by using the well-studied stochastic Ginzburg-Landau equation.~\cite{Hohenberg:RMP:1977} The relevance of Brownian-motion for the parametric spin waves also suggests why it is inherently difficult to reproduce narrow linewidths by straightforward numerical time-integration of magnetization dynamics. In all time-integration schemes, to the best of our knowledge, the effect of finite temperatures is accounted for by including random noise terms to the equations of motion of magnetization, as we have done. While one could expect amplitude fluctuations to reach dynamic equilibrium in a time of $\zk^{-1}$ (e.g., for $\zz = 1.25$, $\zk^{-1} \simeq$ 5 ns for the nanopillars in Ref.~\onlinecite{Mistral:2006}), the ensuing phase fluctuations would require integration times to at least the order of $1/\zD f \simeq$ 0.1 $\zm$s for meaningful comparisons with experiment. Furthermore, it is  necessary to average over a large number of realizations of the stochastic force to recover the appropriate statistics. This represents a computationally-intensive task, so any lighthearted venture into such a program of study will necessarily lead to inconsistent results.

Within the framework of the stochastic model presented here, spectral linewidths of current-driven magnetization oscillations are due to phase noise that is determined purely by the spin-wave intensity and temperature, where the latter controls the level of thermal noise due originating from thermal magnons present in the system. This relationship holds independently of any details concerning the spin-transfer efficiency of the heterostructure $\zs$, notably its magnitude and angular dependence. As such, no attempt is made to discuss absolute linewidths in terms of any external control parameter such as the applied electric current. Furthermore, we expect that the excited spin-wave mode will not be stable to all levels of current-driving, and therefore details concerning the nonlinear instabilities would be required to give any quantitative predictions on the minimum linewidth attainable for a particular system.

In the discussion of frequency modulations due to external sources, we evoked briefly the random fields (\ref{eq:randomfield}) appropriate for the superparamagnetic fluctuations of a single-domain particle.~\cite{Brown:PR:1963} Indeed, transverse components of this random field can also give rise to a stochastic term of the form $f_0(t)$ in (\ref{eq:sto_stochastic}), where only coupling to the (quasi-)uniform mode ($k=0$) is possible given that the random field is taken to be uniform across the magnetic particle. As for the frequency modulations, it is easy to show that the transverse components also lead to an amplitude-independent phase noise of the order of 0.1 MHz. This does not appear to be the dominant mechanism for spin-valve nanopillars at least, where a strong $1/d$-dependence is seen.

In summary, we have developed a stochastic theory of spin-oscillator linewidths based on spin-wave theory. We have derived linear Langevin equations for fluctuations about dynamic equilibrium, which are shown to lead to spectral linewidths that are inversely proportional to the spin-wave intensity. Lower bounds for the linewidth are shown to arise from random fields leading to frequency modulations. These results are shown to give good quantitative agreement with experiments on spin-valve nanopillars.

\begin{acknowledgments}
The author would like to thank Q. Mistral, T. Devolder, P. Crozat, C. Chappert, and T. Schrefl for stimulating discussions. This work was supported by the European Communities programmes ``Human Potential" under Contract No. HRPN-CT-2002-00318 ULTRASWITCH and IST under Contract No. IST-016939 TUNAMOS.
\end{acknowledgments}

% Produces the bibliography via BibTeX.
%\newpage
\bibliography{articles}

\begin{thebibliography}{48}
\expandafter\ifx\csname natexlab\endcsname\relax\def\natexlab#1{#1}\fi
\expandafter\ifx\csname bibnamefont\endcsname\relax
  \def\bibnamefont#1{#1}\fi
\expandafter\ifx\csname bibfnamefont\endcsname\relax
  \def\bibfnamefont#1{#1}\fi
\expandafter\ifx\csname citenamefont\endcsname\relax
  \def\citenamefont#1{#1}\fi
\expandafter\ifx\csname url\endcsname\relax
  \def\url#1{\texttt{#1}}\fi
\expandafter\ifx\csname urlprefix\endcsname\relax\def\urlprefix{URL }\fi
\providecommand{\bibinfo}[2]{#2}
\providecommand{\eprint}[2][]{\url{#2}}

\bibitem[{\citenamefont{Berger}(1996)}]{Berger:PRB:1996}
\bibinfo{author}{\bibfnamefont{L.}~\bibnamefont{Berger}},
  \bibinfo{journal}{Phys. Rev. B} \textbf{\bibinfo{volume}{54}},
  \bibinfo{pages}{9353} (\bibinfo{year}{1996}).

\bibitem[{\citenamefont{Slonczewski}(1996)}]{Slonczewski:JMMM:1996}
\bibinfo{author}{\bibfnamefont{J.~C.} \bibnamefont{Slonczewski}},
  \bibinfo{journal}{J. Magn. Magn. Mater.} \textbf{\bibinfo{volume}{159}},
  \bibinfo{pages}{L1} (\bibinfo{year}{1996}).

\bibitem[{\citenamefont{Slonczewski}(1999)}]{Slonczewski:JMMM:1999}
\bibinfo{author}{\bibfnamefont{J.~C.} \bibnamefont{Slonczewski}},
  \bibinfo{journal}{J. Magn. Magn. Mater.} \textbf{\bibinfo{volume}{195}},
  \bibinfo{pages}{L261} (\bibinfo{year}{1999}).

\bibitem[{\citenamefont{Tsoi et~al.}(1998)\citenamefont{Tsoi, Jansen, Bass,
  Chiang, Seck, Tsoi, and Wyder}}]{Tsoi:PRL:1998}
\bibinfo{author}{\bibfnamefont{M.}~\bibnamefont{Tsoi}},
  \bibinfo{author}{\bibfnamefont{A.~G.~M.} \bibnamefont{Jansen}},
  \bibinfo{author}{\bibfnamefont{J.}~\bibnamefont{Bass}},
  \bibinfo{author}{\bibfnamefont{W.-C.} \bibnamefont{Chiang}},
  \bibinfo{author}{\bibfnamefont{M.}~\bibnamefont{Seck}},
  \bibinfo{author}{\bibfnamefont{V.}~\bibnamefont{Tsoi}}, \bibnamefont{and}
  \bibinfo{author}{\bibfnamefont{P.}~\bibnamefont{Wyder}},
  \bibinfo{journal}{Phys. Rev. Lett.} \textbf{\bibinfo{volume}{80}},
  \bibinfo{pages}{4281} (\bibinfo{year}{1998}).

\bibitem[{\citenamefont{Tsoi et~al.}(2002)\citenamefont{Tsoi, Tsoi, Bass,
  Jansen, and Wyder}}]{Tsoi:PRL:2002}
\bibinfo{author}{\bibfnamefont{M.}~\bibnamefont{Tsoi}},
  \bibinfo{author}{\bibfnamefont{V.}~\bibnamefont{Tsoi}},
  \bibinfo{author}{\bibfnamefont{J.}~\bibnamefont{Bass}},
  \bibinfo{author}{\bibfnamefont{A.~G.~M.} \bibnamefont{Jansen}},
  \bibnamefont{and} \bibinfo{author}{\bibfnamefont{P.}~\bibnamefont{Wyder}},
  \bibinfo{journal}{Phys. Rev. Lett.} \textbf{\bibinfo{volume}{89}},
  \bibinfo{pages}{246803} (\bibinfo{year}{2002}).

\bibitem[{\citenamefont{Rippard et~al.}(2003)\citenamefont{Rippard, Pufall, and
  Silva}}]{Rippard:APL:2003}
\bibinfo{author}{\bibfnamefont{W.~H.} \bibnamefont{Rippard}},
  \bibinfo{author}{\bibfnamefont{M.~R.} \bibnamefont{Pufall}},
  \bibnamefont{and} \bibinfo{author}{\bibfnamefont{T.~J.} \bibnamefont{Silva}},
  \bibinfo{journal}{Appl. Phys. Lett.} \textbf{\bibinfo{volume}{82}},
  \bibinfo{pages}{1260} (\bibinfo{year}{2003}).

\bibitem[{\citenamefont{Kiselev et~al.}(2003)\citenamefont{Kiselev, Sankey,
  Krivorotov, Emley, Schoelkopf, Buhrman, and Ralph}}]{Kiselev:Nature:2003}
\bibinfo{author}{\bibfnamefont{S.~I.} \bibnamefont{Kiselev}},
  \bibinfo{author}{\bibfnamefont{J.~C.} \bibnamefont{Sankey}},
  \bibinfo{author}{\bibfnamefont{I.~N.} \bibnamefont{Krivorotov}},
  \bibinfo{author}{\bibfnamefont{N.~C.} \bibnamefont{Emley}},
  \bibinfo{author}{\bibfnamefont{R.~J.} \bibnamefont{Schoelkopf}},
  \bibinfo{author}{\bibfnamefont{R.~A.} \bibnamefont{Buhrman}},
  \bibnamefont{and} \bibinfo{author}{\bibfnamefont{D.~C.} \bibnamefont{Ralph}},
  \bibinfo{journal}{Nature} \textbf{\bibinfo{volume}{425}},
  \bibinfo{pages}{380} (\bibinfo{year}{2003}).

\bibitem[{\citenamefont{Rippard
  et~al.}(2004{\natexlab{a}})\citenamefont{Rippard, Pufall, Kaka, Russek, and
  Silva}}]{Rippard:PRL:2004}
\bibinfo{author}{\bibfnamefont{W.~H.} \bibnamefont{Rippard}},
  \bibinfo{author}{\bibfnamefont{M.~R.} \bibnamefont{Pufall}},
  \bibinfo{author}{\bibfnamefont{S.}~\bibnamefont{Kaka}},
  \bibinfo{author}{\bibfnamefont{S.~E.} \bibnamefont{Russek}},
  \bibnamefont{and} \bibinfo{author}{\bibfnamefont{T.~J.} \bibnamefont{Silva}},
  \bibinfo{journal}{Phys. Rev. Lett.} \textbf{\bibinfo{volume}{92}},
  \bibinfo{eid}{027201} (\bibinfo{year}{2004}{\natexlab{a}}).

\bibitem[{\citenamefont{Tsoi et~al.}(2004{\natexlab{a}})\citenamefont{Tsoi,
  Sun, Rooks, Koch, and Parkin}}]{Tsoi:PRB:2004}
\bibinfo{author}{\bibfnamefont{M.}~\bibnamefont{Tsoi}},
  \bibinfo{author}{\bibfnamefont{J.~Z.} \bibnamefont{Sun}},
  \bibinfo{author}{\bibfnamefont{M.~J.} \bibnamefont{Rooks}},
  \bibinfo{author}{\bibfnamefont{R.~H.} \bibnamefont{Koch}}, \bibnamefont{and}
  \bibinfo{author}{\bibfnamefont{S.~S.~P.} \bibnamefont{Parkin}},
  \bibinfo{journal}{Phys. Rev. B} \textbf{\bibinfo{volume}{69}},
  \bibinfo{eid}{100406(R)} (\bibinfo{year}{2004}{\natexlab{a}}).

\bibitem[{\citenamefont{Kiselev et~al.}(2004)\citenamefont{Kiselev, Sankey,
  Krivorotov, Emley, Rinkoski, Perez, Buhrman, and Ralph}}]{Kiselev:PRL:2004}
\bibinfo{author}{\bibfnamefont{S.~I.} \bibnamefont{Kiselev}},
  \bibinfo{author}{\bibfnamefont{J.~C.} \bibnamefont{Sankey}},
  \bibinfo{author}{\bibfnamefont{I.~N.} \bibnamefont{Krivorotov}},
  \bibinfo{author}{\bibfnamefont{N.~C.} \bibnamefont{Emley}},
  \bibinfo{author}{\bibfnamefont{M.}~\bibnamefont{Rinkoski}},
  \bibinfo{author}{\bibfnamefont{C.}~\bibnamefont{Perez}},
  \bibinfo{author}{\bibfnamefont{R.~A.} \bibnamefont{Buhrman}},
  \bibnamefont{and} \bibinfo{author}{\bibfnamefont{D.~C.} \bibnamefont{Ralph}},
  \bibinfo{journal}{Phys. Rev. Lett.} \textbf{\bibinfo{volume}{93}},
  \bibinfo{eid}{036601} (\bibinfo{year}{2004}).

\bibitem[{\citenamefont{Covington et~al.}(2004)\citenamefont{Covington,
  AlHajDarwish, Ding, Gokemeijer, and Seigler}}]{Covington:PRB:2004}
\bibinfo{author}{\bibfnamefont{M.}~\bibnamefont{Covington}},
  \bibinfo{author}{\bibfnamefont{M.}~\bibnamefont{AlHajDarwish}},
  \bibinfo{author}{\bibfnamefont{Y.}~\bibnamefont{Ding}},
  \bibinfo{author}{\bibfnamefont{N.~J.} \bibnamefont{Gokemeijer}},
  \bibnamefont{and} \bibinfo{author}{\bibfnamefont{M.~A.}
  \bibnamefont{Seigler}}, \bibinfo{journal}{Phys. Rev. B}
  \textbf{\bibinfo{volume}{69}}, \bibinfo{eid}{184406} (\bibinfo{year}{2004}).

\bibitem[{\citenamefont{Tsoi et~al.}(2004{\natexlab{b}})\citenamefont{Tsoi,
  Tsoi, and Wyder}}]{Tsoi:PRB:2004b}
\bibinfo{author}{\bibfnamefont{M.}~\bibnamefont{Tsoi}},
  \bibinfo{author}{\bibfnamefont{V.~S.} \bibnamefont{Tsoi}}, \bibnamefont{and}
  \bibinfo{author}{\bibfnamefont{P.}~\bibnamefont{Wyder}},
  \bibinfo{journal}{Phys. Rev. B} \textbf{\bibinfo{volume}{70}},
  \bibinfo{eid}{012405} (\bibinfo{year}{2004}{\natexlab{b}}).

\bibitem[{\citenamefont{Tsoi et~al.}(2004{\natexlab{c}})\citenamefont{Tsoi,
  Sun, and Parkin}}]{Tsoi:PRL:2004}
\bibinfo{author}{\bibfnamefont{M.}~\bibnamefont{Tsoi}},
  \bibinfo{author}{\bibfnamefont{J.~Z.} \bibnamefont{Sun}}, \bibnamefont{and}
  \bibinfo{author}{\bibfnamefont{S.~S.~P.} \bibnamefont{Parkin}},
  \bibinfo{journal}{Phys. Rev. Lett.} \textbf{\bibinfo{volume}{93}},
  \bibinfo{eid}{036602} (\bibinfo{year}{2004}{\natexlab{c}}).

\bibitem[{\citenamefont{Rippard
  et~al.}(2004{\natexlab{b}})\citenamefont{Rippard, Pufall, Kaka, Silva, and
  Russek}}]{Rippard:PRB:2004}
\bibinfo{author}{\bibfnamefont{W.~H.} \bibnamefont{Rippard}},
  \bibinfo{author}{\bibfnamefont{M.~R.} \bibnamefont{Pufall}},
  \bibinfo{author}{\bibfnamefont{S.}~\bibnamefont{Kaka}},
  \bibinfo{author}{\bibfnamefont{T.~J.} \bibnamefont{Silva}}, \bibnamefont{and}
  \bibinfo{author}{\bibfnamefont{S.~E.} \bibnamefont{Russek}},
  \bibinfo{journal}{Phys. Rev. B} \textbf{\bibinfo{volume}{70}},
  \bibinfo{eid}{100406(R)} (\bibinfo{year}{2004}{\natexlab{b}}).

\bibitem[{\citenamefont{Krivorotov et~al.}(2005)\citenamefont{Krivorotov,
  Emley, Sankey, Kiselev, Ralph, and Buhrman}}]{Krivorotov:Science:2005}
\bibinfo{author}{\bibfnamefont{I.~N.} \bibnamefont{Krivorotov}},
  \bibinfo{author}{\bibfnamefont{N.~C.} \bibnamefont{Emley}},
  \bibinfo{author}{\bibfnamefont{J.~C.} \bibnamefont{Sankey}},
  \bibinfo{author}{\bibfnamefont{S.~I.} \bibnamefont{Kiselev}},
  \bibinfo{author}{\bibfnamefont{D.~C.} \bibnamefont{Ralph}}, \bibnamefont{and}
  \bibinfo{author}{\bibfnamefont{R.~A.} \bibnamefont{Buhrman}},
  \bibinfo{journal}{Science} \textbf{\bibinfo{volume}{307}},
  \bibinfo{pages}{228} (\bibinfo{year}{2005}).

\bibitem[{\citenamefont{Kiselev et~al.}(2005)\citenamefont{Kiselev, Sankey,
  Krivorotov, Emley, Garcia, Buhrman, and Ralph}}]{Kiselev:PRB:2005}
\bibinfo{author}{\bibfnamefont{S.~I.} \bibnamefont{Kiselev}},
  \bibinfo{author}{\bibfnamefont{J.~C.} \bibnamefont{Sankey}},
  \bibinfo{author}{\bibfnamefont{I.~N.} \bibnamefont{Krivorotov}},
  \bibinfo{author}{\bibfnamefont{N.~C.} \bibnamefont{Emley}},
  \bibinfo{author}{\bibfnamefont{A.~G.~F.} \bibnamefont{Garcia}},
  \bibinfo{author}{\bibfnamefont{R.~A.} \bibnamefont{Buhrman}},
  \bibnamefont{and} \bibinfo{author}{\bibfnamefont{D.~C.} \bibnamefont{Ralph}},
  \bibinfo{journal}{Phys. Rev. B} \textbf{\bibinfo{volume}{72}},
  \bibinfo{eid}{064430} (\bibinfo{year}{2005}).

\bibitem[{\citenamefont{Kaka et~al.}(2005)\citenamefont{Kaka, Pufall, Rippard,
  Silva, Russek, and Katine}}]{Kaka:Nature:2005}
\bibinfo{author}{\bibfnamefont{S.}~\bibnamefont{Kaka}},
  \bibinfo{author}{\bibfnamefont{M.~R.} \bibnamefont{Pufall}},
  \bibinfo{author}{\bibfnamefont{W.~H.} \bibnamefont{Rippard}},
  \bibinfo{author}{\bibfnamefont{T.~J.} \bibnamefont{Silva}},
  \bibinfo{author}{\bibfnamefont{S.~E.} \bibnamefont{Russek}},
  \bibnamefont{and} \bibinfo{author}{\bibfnamefont{J.~A.}
  \bibnamefont{Katine}}, \bibinfo{journal}{Nature}
  \textbf{\bibinfo{volume}{437}}, \bibinfo{pages}{389} (\bibinfo{year}{2005}).

\bibitem[{\citenamefont{Mancoff et~al.}(2005)\citenamefont{Mancoff, Rizzo,
  Engel, and Tehrani}}]{Mancoff:Nature:2005}
\bibinfo{author}{\bibfnamefont{F.~B.} \bibnamefont{Mancoff}},
  \bibinfo{author}{\bibfnamefont{N.~D.} \bibnamefont{Rizzo}},
  \bibinfo{author}{\bibfnamefont{B.~N.} \bibnamefont{Engel}}, \bibnamefont{and}
  \bibinfo{author}{\bibfnamefont{S.}~\bibnamefont{Tehrani}},
  \bibinfo{journal}{Nature} \textbf{\bibinfo{volume}{437}},
  \bibinfo{pages}{393} (\bibinfo{year}{2005}).

\bibitem[{\citenamefont{Mistral et~al.}(2006)\citenamefont{Mistral, Deac,
  Grollier, Redon, Liu, Li, Wang, Dieny, and Devolder}}]{Mistral:MSEB:2006}
\bibinfo{author}{\bibfnamefont{Q.}~\bibnamefont{Mistral}},
  \bibinfo{author}{\bibfnamefont{A.}~\bibnamefont{Deac}},
  \bibinfo{author}{\bibfnamefont{J.}~\bibnamefont{Grollier}},
  \bibinfo{author}{\bibfnamefont{O.}~\bibnamefont{Redon}},
  \bibinfo{author}{\bibfnamefont{Y.}~\bibnamefont{Liu}},
  \bibinfo{author}{\bibfnamefont{M.}~\bibnamefont{Li}},
  \bibinfo{author}{\bibfnamefont{P.}~\bibnamefont{Wang}},
  \bibinfo{author}{\bibfnamefont{B.}~\bibnamefont{Dieny}}, \bibnamefont{and}
  \bibinfo{author}{\bibfnamefont{T.}~\bibnamefont{Devolder}},
  \bibinfo{journal}{Mat. Sci. Eng. B} \textbf{\bibinfo{volume}{126}},
  \bibinfo{pages}{267} (\bibinfo{year}{2006}).

\bibitem[{\citenamefont{Rezende et~al.}(2005)\citenamefont{Rezende, de~Aguiar,
  and Azevedo}}]{Rezende:PRL:2005}
\bibinfo{author}{\bibfnamefont{S.~M.} \bibnamefont{Rezende}},
  \bibinfo{author}{\bibfnamefont{F.~M.} \bibnamefont{de~Aguiar}},
  \bibnamefont{and} \bibinfo{author}{\bibfnamefont{A.}~\bibnamefont{Azevedo}},
  \bibinfo{journal}{Phys. Rev. Lett.} \textbf{\bibinfo{volume}{94}},
  \bibinfo{eid}{037202} (\bibinfo{year}{2005}).

\bibitem[{\citenamefont{Slavin and Kabos}(2005)}]{Slavin:IEEE:2005}
\bibinfo{author}{\bibfnamefont{A.}~\bibnamefont{Slavin}} \bibnamefont{and}
  \bibinfo{author}{\bibfnamefont{P.}~\bibnamefont{Kabos}},
  \bibinfo{journal}{IEEE Trans. Magn.} \textbf{\bibinfo{volume}{41}},
  \bibinfo{pages}{1264} (\bibinfo{year}{2005}).

\bibitem[{\citenamefont{Lee et~al.}(2004)\citenamefont{Lee, Deac, Redon,
  Nozi{\`e}res, and Dieny}}]{Lee:NM:2004}
\bibinfo{author}{\bibfnamefont{K.-J.} \bibnamefont{Lee}},
  \bibinfo{author}{\bibfnamefont{A.}~\bibnamefont{Deac}},
  \bibinfo{author}{\bibfnamefont{O.}~\bibnamefont{Redon}},
  \bibinfo{author}{\bibfnamefont{J.-P.} \bibnamefont{Nozi{\`e}res}},
  \bibnamefont{and} \bibinfo{author}{\bibfnamefont{B.}~\bibnamefont{Dieny}},
  \bibinfo{journal}{Nature Mat.} \textbf{\bibinfo{volume}{3}},
  \bibinfo{pages}{877} (\bibinfo{year}{2004}).

\bibitem[{\citenamefont{Xi and Lin}(2004)}]{Xi:PRB:2004}
\bibinfo{author}{\bibfnamefont{H.}~\bibnamefont{Xi}} \bibnamefont{and}
  \bibinfo{author}{\bibfnamefont{Z.}~\bibnamefont{Lin}},
  \bibinfo{journal}{Phys. Rev. B} \textbf{\bibinfo{volume}{70}},
  \bibinfo{pages}{092403} (\bibinfo{year}{2004}).

\bibitem[{\citenamefont{Russek et~al.}(2005)\citenamefont{Russek, Kaka,
  Rippard, Pufall, and Silva}}]{Russek:PRB:2005}
\bibinfo{author}{\bibfnamefont{S.~E.} \bibnamefont{Russek}},
  \bibinfo{author}{\bibfnamefont{S.}~\bibnamefont{Kaka}},
  \bibinfo{author}{\bibfnamefont{W.~H.} \bibnamefont{Rippard}},
  \bibinfo{author}{\bibfnamefont{M.~R.} \bibnamefont{Pufall}},
  \bibnamefont{and} \bibinfo{author}{\bibfnamefont{T.~J.} \bibnamefont{Silva}},
  \bibinfo{journal}{Phys. Rev. B} \textbf{\bibinfo{volume}{71}},
  \bibinfo{eid}{104425} (\bibinfo{year}{2005}).

\bibitem[{\citenamefont{Bertotti et~al.}(2005)\citenamefont{Bertotti, Serpico,
  Mayergoyz, Magni, d'Aquino, and Bonin}}]{Bertotti:PRL:2005}
\bibinfo{author}{\bibfnamefont{G.}~\bibnamefont{Bertotti}},
  \bibinfo{author}{\bibfnamefont{C.}~\bibnamefont{Serpico}},
  \bibinfo{author}{\bibfnamefont{I.~D.} \bibnamefont{Mayergoyz}},
  \bibinfo{author}{\bibfnamefont{A.}~\bibnamefont{Magni}},
  \bibinfo{author}{\bibfnamefont{M.}~\bibnamefont{d'Aquino}}, \bibnamefont{and}
  \bibinfo{author}{\bibfnamefont{R.}~\bibnamefont{Bonin}},
  \bibinfo{journal}{Phys. Rev. Lett.} \textbf{\bibinfo{volume}{94}},
  \bibinfo{pages}{127206} (\bibinfo{year}{2005}).

\bibitem[{\citenamefont{Montigny and Miltat}(2005)}]{Montigny:JAP:2005}
\bibinfo{author}{\bibfnamefont{B.}~\bibnamefont{Montigny}} \bibnamefont{and}
  \bibinfo{author}{\bibfnamefont{J.}~\bibnamefont{Miltat}},
  \bibinfo{journal}{J. Appl. Phys.} \textbf{\bibinfo{volume}{97}},
  \bibinfo{eid}{10C708} (\bibinfo{year}{2005}).

\bibitem[{\citenamefont{Xiao et~al.}(2005)\citenamefont{Xiao, Zangwill, and
  Stiles}}]{Xiao:PRB:2005}
\bibinfo{author}{\bibfnamefont{J.}~\bibnamefont{Xiao}},
  \bibinfo{author}{\bibfnamefont{A.}~\bibnamefont{Zangwill}}, \bibnamefont{and}
  \bibinfo{author}{\bibfnamefont{M.~D.} \bibnamefont{Stiles}},
  \bibinfo{journal}{Phys. Rev. B} \textbf{\bibinfo{volume}{72}},
  \bibinfo{eid}{014446} (\bibinfo{year}{2005}).

\bibitem[{\citenamefont{Mistral et~al.}()\citenamefont{Mistral, Kim, Devolder,
  Crozat, Chappert, Katine, Carey, and Ito}}]{Mistral:2006}
\bibinfo{author}{\bibfnamefont{Q.}~\bibnamefont{Mistral}},
  \bibinfo{author}{\bibfnamefont{J.-V.} \bibnamefont{Kim}},
  \bibinfo{author}{\bibfnamefont{T.}~\bibnamefont{Devolder}},
  \bibinfo{author}{\bibfnamefont{P.}~\bibnamefont{Crozat}},
  \bibinfo{author}{\bibfnamefont{C.}~\bibnamefont{Chappert}},
  \bibinfo{author}{\bibfnamefont{J.~A.} \bibnamefont{Katine}},
  \bibinfo{author}{\bibfnamefont{M.~J.} \bibnamefont{Carey}}, \bibnamefont{and}
  \bibinfo{author}{\bibfnamefont{K.}~\bibnamefont{Ito}}, \bibinfo{note}{{Appl.
  Phys. Lett., to appear}}.

\bibitem[{\citenamefont{Morgenthaler}(1960)}]{Morgenthaler:JAP:1960}
\bibinfo{author}{\bibfnamefont{F.~R.} \bibnamefont{Morgenthaler}},
  \bibinfo{journal}{J. Appl. Phys.} \textbf{\bibinfo{volume}{31}},
  \bibinfo{pages}{S95} (\bibinfo{year}{1960}).

\bibitem[{\citenamefont{Schlomann et~al.}(1960)\citenamefont{Schlomann, Green,
  and Milano}}]{Schlomann:JAP:1960}
\bibinfo{author}{\bibfnamefont{E.}~\bibnamefont{Schlomann}},
  \bibinfo{author}{\bibfnamefont{J.~J.} \bibnamefont{Green}}, \bibnamefont{and}
  \bibinfo{author}{\bibfnamefont{U.}~\bibnamefont{Milano}},
  \bibinfo{journal}{J. Appl. Phys.} \textbf{\bibinfo{volume}{31}},
  \bibinfo{pages}{S386} (\bibinfo{year}{1960}).

\bibitem[{\citenamefont{{L'vov}}(1994)}]{Lvov:1994}
\bibinfo{author}{\bibfnamefont{V.}~\bibnamefont{{L'vov}}},
  \emph{\bibinfo{title}{Wave Turbulence Under Parametric Excitation}}
  (\bibinfo{publisher}{Springer-Verlag}, \bibinfo{address}{Berlin},
  \bibinfo{year}{1994}).

\bibitem[{\citenamefont{Slavin et~al.}(1994)\citenamefont{Slavin, Kalinikos,
  and Korshikov}}]{Slavin:1994}
\bibinfo{author}{\bibfnamefont{A.~N.} \bibnamefont{Slavin}},
  \bibinfo{author}{\bibfnamefont{B.~A.} \bibnamefont{Kalinikos}},
  \bibnamefont{and} \bibinfo{author}{\bibfnamefont{N.~G.}
  \bibnamefont{Korshikov}}, in \emph{\bibinfo{booktitle}{Nonlinear phenomena
  and chaos in magnetic materials}}, edited by
  \bibinfo{editor}{\bibfnamefont{P.~E.} \bibnamefont{Wigen}}
  (\bibinfo{publisher}{World Scientific}, \bibinfo{year}{1994}), pp.
  \bibinfo{pages}{209--248}.

\bibitem[{\citenamefont{Giovannini et~al.}(2004)\citenamefont{Giovannini,
  Montoncello, Nizzoli, Gubbiotti, Carlotti, Okuno, Shinjo, and
  Grimsditch}}]{Giovannini:PRB:2004}
\bibinfo{author}{\bibfnamefont{L.}~\bibnamefont{Giovannini}},
  \bibinfo{author}{\bibfnamefont{F.}~\bibnamefont{Montoncello}},
  \bibinfo{author}{\bibfnamefont{F.}~\bibnamefont{Nizzoli}},
  \bibinfo{author}{\bibfnamefont{G.}~\bibnamefont{Gubbiotti}},
  \bibinfo{author}{\bibfnamefont{G.}~\bibnamefont{Carlotti}},
  \bibinfo{author}{\bibfnamefont{T.}~\bibnamefont{Okuno}},
  \bibinfo{author}{\bibfnamefont{T.}~\bibnamefont{Shinjo}}, \bibnamefont{and}
  \bibinfo{author}{\bibfnamefont{M.}~\bibnamefont{Grimsditch}},
  \bibinfo{journal}{Phys. Rev. B} \textbf{\bibinfo{volume}{70}},
  \bibinfo{eid}{172404} (\bibinfo{year}{2004}).

\bibitem[{\citenamefont{Grimsditch et~al.}(2004)\citenamefont{Grimsditch,
  Giovannini, Montoncello, Nizzoli, Leaf, and Kaper}}]{Grimsditch:PRB:2004}
\bibinfo{author}{\bibfnamefont{M.}~\bibnamefont{Grimsditch}},
  \bibinfo{author}{\bibfnamefont{L.}~\bibnamefont{Giovannini}},
  \bibinfo{author}{\bibfnamefont{F.}~\bibnamefont{Montoncello}},
  \bibinfo{author}{\bibfnamefont{F.}~\bibnamefont{Nizzoli}},
  \bibinfo{author}{\bibfnamefont{G.~K.} \bibnamefont{Leaf}}, \bibnamefont{and}
  \bibinfo{author}{\bibfnamefont{H.~G.} \bibnamefont{Kaper}},
  \bibinfo{journal}{Phys. Rev. B} \textbf{\bibinfo{volume}{70}},
  \bibinfo{eid}{054409} (\bibinfo{year}{2004}).

\bibitem[{\citenamefont{McMichael and Stiles}(2005)}]{McMichael:JAP:2005}
\bibinfo{author}{\bibfnamefont{R.~D.} \bibnamefont{McMichael}}
  \bibnamefont{and} \bibinfo{author}{\bibfnamefont{M.~D.}
  \bibnamefont{Stiles}}, \bibinfo{journal}{J. Appl. Phys.}
  \textbf{\bibinfo{volume}{97}}, \bibinfo{eid}{10J901} (\bibinfo{year}{2005}).

\bibitem[{\citenamefont{Guslienko et~al.}(2003)\citenamefont{Guslienko,
  Chantrell, and Slavin}}]{Guslienko:PRB:2003}
\bibinfo{author}{\bibfnamefont{K.~Y.} \bibnamefont{Guslienko}},
  \bibinfo{author}{\bibfnamefont{R.~W.} \bibnamefont{Chantrell}},
  \bibnamefont{and} \bibinfo{author}{\bibfnamefont{A.~N.}
  \bibnamefont{Slavin}}, \bibinfo{journal}{Phys. Rev. B}
  \textbf{\bibinfo{volume}{68}}, \bibinfo{eid}{024422} (\bibinfo{year}{2003}).

\bibitem[{\citenamefont{Guslienko and Slavin}(2005)}]{Guslienko:PRB:2005}
\bibinfo{author}{\bibfnamefont{K.~Y.} \bibnamefont{Guslienko}}
  \bibnamefont{and} \bibinfo{author}{\bibfnamefont{A.~N.}
  \bibnamefont{Slavin}}, \bibinfo{journal}{Phys. Rev. B}
  \textbf{\bibinfo{volume}{72}}, \bibinfo{eid}{014463} (\bibinfo{year}{2005}).

\bibitem[{\citenamefont{Bayer et~al.}(2005)\citenamefont{Bayer, Jorzick,
  Hillebrands, Demokritov, Kouba, Bozinoski, Slavin, Guslienko, Berkov, Gorn
  et~al.}}]{Bayer:PRB:2005}
\bibinfo{author}{\bibfnamefont{C.}~\bibnamefont{Bayer}},
  \bibinfo{author}{\bibfnamefont{J.}~\bibnamefont{Jorzick}},
  \bibinfo{author}{\bibfnamefont{B.}~\bibnamefont{Hillebrands}},
  \bibinfo{author}{\bibfnamefont{S.~O.} \bibnamefont{Demokritov}},
  \bibinfo{author}{\bibfnamefont{R.}~\bibnamefont{Kouba}},
  \bibinfo{author}{\bibfnamefont{R.}~\bibnamefont{Bozinoski}},
  \bibinfo{author}{\bibfnamefont{A.~N.} \bibnamefont{Slavin}},
  \bibinfo{author}{\bibfnamefont{K.~Y.} \bibnamefont{Guslienko}},
  \bibinfo{author}{\bibfnamefont{D.~V.} \bibnamefont{Berkov}},
  \bibinfo{author}{\bibfnamefont{N.~L.} \bibnamefont{Gorn}},
  \bibnamefont{et~al.}, \bibinfo{journal}{Phys. Rev. B}
  \textbf{\bibinfo{volume}{72}}, \bibinfo{eid}{064427} (\bibinfo{year}{2005}).

\bibitem[{\citenamefont{Mikhailov}(1976)}]{Mikhailov:JETP:1976}
\bibinfo{author}{\bibfnamefont{A.~S.} \bibnamefont{Mikhailov}},
  \bibinfo{journal}{Sov. Phys. JETP} \textbf{\bibinfo{volume}{42}},
  \bibinfo{pages}{267} (\bibinfo{year}{1976}).

\bibitem[{\citenamefont{Zakharov and {L'vov}}(1971)}]{Zakharov:JETP:1971}
\bibinfo{author}{\bibfnamefont{V.~E.} \bibnamefont{Zakharov}} \bibnamefont{and}
  \bibinfo{author}{\bibfnamefont{V.~S.} \bibnamefont{{L'vov}}},
  \bibinfo{journal}{Sov. Phys. JETP} \textbf{\bibinfo{volume}{33}},
  \bibinfo{pages}{1113} (\bibinfo{year}{1971}).

\bibitem[{\citenamefont{Kubo}(1962)}]{Kubo:1962}
\bibinfo{author}{\bibfnamefont{R.}~\bibnamefont{Kubo}}, in
  \emph{\bibinfo{booktitle}{Fluctuation, relaxation and resonance in magnetic
  systems}}, edited by \bibinfo{editor}{\bibfnamefont{D.}~\bibnamefont{{Ter
  Haar}}} (\bibinfo{publisher}{Oliver \& Boyd}, \bibinfo{year}{1962}), pp.
  \bibinfo{pages}{23--68}.

\bibitem[{\citenamefont{Risken}(1989)}]{Risken:1989}
\bibinfo{author}{\bibfnamefont{H.}~\bibnamefont{Risken}},
  \emph{\bibinfo{title}{The Fokker-Planck Equation}}
  (\bibinfo{publisher}{Springer-Verlag}, \bibinfo{address}{Berlin},
  \bibinfo{year}{1989}).

\bibitem[{\citenamefont{{Brown, Jr.}}(1963)}]{Brown:PR:1963}
\bibinfo{author}{\bibfnamefont{W.~F.} \bibnamefont{{Brown, Jr.}}},
  \bibinfo{journal}{Phys. Rev.} \textbf{\bibinfo{volume}{130}},
  \bibinfo{pages}{1677} (\bibinfo{year}{1963}).

\bibitem[{\citenamefont{Foros et~al.}(2005)\citenamefont{Foros, Brataas,
  Tserkovnyak, and Bauer}}]{Foros:PRL:2005}
\bibinfo{author}{\bibfnamefont{J.}~\bibnamefont{Foros}},
  \bibinfo{author}{\bibfnamefont{A.}~\bibnamefont{Brataas}},
  \bibinfo{author}{\bibfnamefont{Y.}~\bibnamefont{Tserkovnyak}},
  \bibnamefont{and} \bibinfo{author}{\bibfnamefont{G.~E.~W.}
  \bibnamefont{Bauer}}, \bibinfo{journal}{Phys. Rev. Lett.}
  \textbf{\bibinfo{volume}{95}}, \bibinfo{pages}{016601}
  (\bibinfo{year}{2005}).

\bibitem[{\citenamefont{Sankey et~al.}(2005)\citenamefont{Sankey, Krivorotov,
  Kiselev, Braganca, Emley, Buhrman, and Ralph}}]{Sankey:PRB:2005}
\bibinfo{author}{\bibfnamefont{J.~C.} \bibnamefont{Sankey}},
  \bibinfo{author}{\bibfnamefont{I.~N.} \bibnamefont{Krivorotov}},
  \bibinfo{author}{\bibfnamefont{S.~I.} \bibnamefont{Kiselev}},
  \bibinfo{author}{\bibfnamefont{P.~M.} \bibnamefont{Braganca}},
  \bibinfo{author}{\bibfnamefont{N.~C.} \bibnamefont{Emley}},
  \bibinfo{author}{\bibfnamefont{R.~A.} \bibnamefont{Buhrman}},
  \bibnamefont{and} \bibinfo{author}{\bibfnamefont{D.~C.} \bibnamefont{Ralph}},
  \bibinfo{journal}{Phys. Rev. B} \textbf{\bibinfo{volume}{72}},
  \bibinfo{eid}{224427} (\bibinfo{year}{2005}).

\bibitem[{\citenamefont{Xiao et~al.}(2004)\citenamefont{Xiao, Zangwill, and
  Stiles}}]{Xiao:PRB:2004}
\bibinfo{author}{\bibfnamefont{J.}~\bibnamefont{Xiao}},
  \bibinfo{author}{\bibfnamefont{A.}~\bibnamefont{Zangwill}}, \bibnamefont{and}
  \bibinfo{author}{\bibfnamefont{M.~D.} \bibnamefont{Stiles}},
  \bibinfo{journal}{Phys. Rev. B} \textbf{\bibinfo{volume}{70}},
  \bibinfo{eid}{172405} (\bibinfo{year}{2004}).

\bibitem[{\citenamefont{Slavin and Tiberkevich}(2005)}]{Slavin:PRL:2005}
\bibinfo{author}{\bibfnamefont{A.}~\bibnamefont{Slavin}} \bibnamefont{and}
  \bibinfo{author}{\bibfnamefont{V.}~\bibnamefont{Tiberkevich}},
  \bibinfo{journal}{Phys. Rev. Lett.} \textbf{\bibinfo{volume}{95}},
  \bibinfo{eid}{237201} (\bibinfo{year}{2005}).

\bibitem[{\citenamefont{Hohenberg and Halperin}(1977)}]{Hohenberg:RMP:1977}
\bibinfo{author}{\bibfnamefont{P.~C.} \bibnamefont{Hohenberg}}
  \bibnamefont{and} \bibinfo{author}{\bibfnamefont{B.~I.}
  \bibnamefont{Halperin}}, \bibinfo{journal}{Rev. Mod. Phys.}
  \textbf{\bibinfo{volume}{49}}, \bibinfo{pages}{435} (\bibinfo{year}{1977}).

\end{thebibliography}

%\newpage
%\printfigures

\end{document}